\documentclass[aps,prl,twocolumn,showpacs]{revtex4}

\usepackage{graphicx}
\begin{document}


\title{Gigantic anisotropic uniaxial pressure effect on superconductivity 
 within the CuO$_{2}$ plane of La$_{1.64}$Eu$_{0.2}$Sr$_{0.16}$CuO$_{4}$ 
 - strain control of stripe criticality}

\author{N. Takeshita$^{1}$, T. Sasagawa$^{2,3}$, T. Sugioka$^{2}$, 
Y. Tokura$^{1,4}$, and H. Takagi$^{1,2,3}$}
\affiliation{$^{1}$ Correlated Electron Research Center (CERC), 
 National Institute of Advanced Industrial Science and Technology (AIST), 
 Tsukuba, Ibaraki 305-8062, Japan}
\affiliation{$^{2}$ Department of Advanced Materials Science, 
 University of Tokyo, 5-1-5 Kashiwa-no-ha, Kashiwa 277-8581, Japan}
\affiliation{$^{3}$ CREST, Japan Science and Technology Corporation (JST)}
\affiliation{$^{4}$ Department of Applied Physics, University of Tokyo, 
 7-3-1 Hongo, Bunkyo-ku, Tokyo 113-8656, Japan}

\date{\today}

\begin{abstract}
The effect of uniaxial pressure on superconductivity was examined for 
a high-{\it T}$_{\rm c}$ cuprate La$_{1.64}$Eu$_{0.2}$Sr$_{0.16}$CuO$_{4}$, 
which is located at the boundary between the superconducting 
and stripe phases. We found remarkably large anisotropy of 
the uniaxial pressure effect not only between the in-plane and 
out-of-plane pressures but also {\it within the CuO$_{2}$-plane}. 
When the pressure is applied along the tetragonal [110] direction, 
we found the largest pressure effect ever observed in cuprates, 
{\it dT$_{\rm c}$/dP}$_{[110]}$ $\sim$ 2.5 K/kbar, 
while the change of {\it T}$_{\rm c}$ was not appreciable 
when applied along [100]. This substantial in-plane anisotropy 
is attributed to an intimate link between the symmetry of 
the one-dimensional stripes and that of the strain produced 
within the CuO$_{2}$ plane.
\end{abstract}

\maketitle

Recently, self-organized states of strongly correlated electrons, 
created by the intimate interplay between charge, spin, and orbital 
degrees of freedom, have been attracting considerable interest, 
because of the rich variety of physics behind it. 
One of the hallmarks of such self-organized states is the ``stripe" 
phase in high-{\it T}$_{\rm c}$ cuprates~\cite{ref1}. 
Stripes in high-{\it T}$_{\rm c}$ cuprates form a periodic pattern of 
charge and spin rivers running along the tetragonal [100] or [010] directions 
(parallel to the Cu-O bonds). 
The signature of stripes is most significant 
in La$_{2}$CuO$_{4}$-based cuprates. 
In the prototypical cuprates a pronounced suppression of superconductivity 
is observed due to the formation of static stripes when the doping level 
is close to a magic number {\it x} = 1/8~\cite{ref2} and/or when 
the crystal has a low temperature tetragonal (LTT) structure~\cite{ref3}. 
The former is attributed to commensurability of the stripe pattern 
with the lattice. The latter is due to a tilting of the CuO$_{6}$ 
octahedron towards [100] in the LTT phase, which stabilizes the formation 
of charge and spin channels running along the [100] direction.

To date, {\it statically} ordered stripes have been observed only 
in La$_{2}$CuO$_{4}$-based cuprates. 
It has been argued, however, that the stripe instability is not specific to 
the La$_{2}$CuO$_{4}$ family but generic to the two-dimensional 
CuO$_{2}$ planes. 
The observation of incommensurate spin correlations in 
YBa$_{2}$Cu$_{3}$O$_{7-\delta}$ by inelastic neutron scattering~\cite{ref4} 
suggests that dynamical stripe fluctuations are present universally 
in layered cuprates, though alternative views have been raised~\cite{ref5}. 
A close link between the stripe formation and high-{\it T}$_{\rm c}$ 
superconductivity has been discussed. 
The static stripes seem to suppress and to compete with high-{\it T}$_{\rm c}$ 
superconductivity. 
We clearly observe an anomalous absence of superconductivity around 
{\it x} = 1/8 in La$_{2-x}$Ba$_{x}$CuO$_{4}$~\cite{ref2}, 
where {\it statically} ordered stripes are formed. 
Dynamically fluctuating stripes, however, have been argued to 
promote superconductivity and are now taken as one of potential candidates 
for the mechanism of high-{\it T}$_{\rm c}$ 
superconductivity~\cite{ref6,ref7,ref8}. 

Such a delicate balance between stripes and high-{\it T}$_{\rm c}$ 
superconductivity has been promoting attempt to finely tune the phase 
competition using external parameters such as pressure~\cite{ref9,ref10} 
and magnetic field~\cite{ref11}. 
Enhancement of superconductivity by {\it hydrostatic} pressure around 
the 1/8 anomaly has been known for many years~\cite{ref9,ref10}. 
In this work, we propose {\it uniaxial} pressure as a novel and promising 
control parameter to explore the critical region of the 
stripe-superconductivity phase boundary, which provides by far 
richer information than the conventional hydrostatic pressure 
and therefore is distinct. 
The stripes in high-{\it T}$_{\rm c}$ cuprates are one-dimensional 
self-organized objects formed within the two-dimensional CuO$_{2}$ plane 
and, therefore, a strong coupling with an anisotropic distortion 
of the tetragonal CuO$_{2}$ plane is anticipated. 
The fact that the stripes are stabilized in the presence of LTT distortion 
along [100] can be understood in terms of such a coupling. 
Hydrostatic pressure is essentially a volume effect and cannot generate 
any anisotropic distortion. 
Uniaxial pressure, on the other hand, produces an anisotropic lattice 
distortion and can thereby change the symmetry of lattice. 
This will modify the stability of stripes drastically and hence 
superconductivity. 
In the past, so-called pressure control of stripes and superconductivity 
has been conducted by utilizing {\it hydrostatic} pressure, 
except for a very limited number of uniaxial studies~\cite{ref12}.

\begin{figure}[b]
\includegraphics[width=0.8\linewidth]{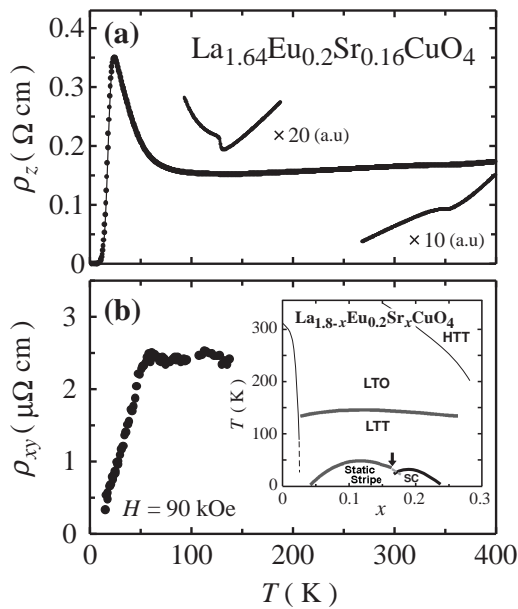}
\caption{(a) Temperature dependence of the out-of-plane resistivity 
 $\rho$$_{z}$ for La$_{1.64}$Eu$_{0.2}$Sr$_{0.16}$CuO$_{4}$ single crystal. 
 Two anomalies associated with the structural phase transitions 
 from high temperature tetragonal (HTT) to 
 low temperature orthorhombic (LTO) and from LTO to 
 low temperature tetragonal (LTT) can be clearly seen in the magnified curves. 
 (b) Temperature dependence of the Hall resistivity $\rho$$_{xy}$, 
 where the anomalous decrease of $\rho$$_{xy}$ at around 50 K marks 
 the onset of the static stripe formation. 
 Inset shows a schematic phase diagram of 
 La$_{1.8-x}$Eu$_{0.2}$Sr$_{x}$CuO$_{4}$.}
\label{fig1}
\end{figure}

We have attempted to examine such a coupling between stripes and 
the anisotropic distortion using La$_{1.8-x}$Eu$_{0.2}$Sr$_{x}$CuO$_{4}$. 
It is well known that the introduction of rare earth ions to the La site 
tends to stabilize the LTT structure~\cite{ref13}. 
In La$_{1.8-x}$Eu$_{0.2}$Sr$_{x}$CuO$_{4}$, by introducing 10\% of Eu 
for La, the LTT phase is stabilized over a wide range of hole concentration 
from {\it x} = 0 to 0.3~\cite{ref14}. 
This provides a unique opportunity to observe the stripe criticality 
in the absence of the first order LTO-LTT phase transition. 
More interestingly, the underdoped side of the superconducting dome 
in the phase diagram gives a way to a ``stripe" magnet, 
as if the superconducting phase were replaced with the ``stripe" 
phase~\cite{ref14}, 
as schematically shown in the inset of Figure~\ref{fig1}(b). 
As a result, an intriguing phase competition between the stripe phase 
and the superconducting phase appears around a critical composition 
of {\it x} = 0.16 which appears as a bicritical point. 
It should be emphasized that the boundary is located away from 
the 1/8 anomaly, which implies that commensurability is less important. 

La$_{1.64}$Eu$_{0.2}$Sr$_{0.16}$CuO$_{4}$ single crystals in critical 
vicinity to the phase boundary were grown by a traveling solvent 
floating zone (TSFZ) technique~\cite{ref15a,ref15b}. 
Three different samples were cut from the same crystalline rod. 
At ambient pressure, all the crystals showed superconducting transition 
at around 10 K. 
The out-of-plane resistivity $\rho$$_{z}$, shown in Figure~\ref{fig1}(a), 
indicates two clear anomalies at 340 K and 125 K. 
These correspond to transitions from HTT to LTO and from LTO to LTT, 
respectively. 
As shown in Fig.~\ref{fig1}(b), the signature of static stripe formation 
was observed below 50 K, which manifests itself as a rapid decrease 
of Hall resistivity $\rho$$_{xy}$ with decreasing temperature~\cite{ref16}. 
Whether this stripe signature represents the coexistence of stripes 
and superconductivity or a two-phase admixture is not clear at this stage, 
but does not affect the main point of this paper. 

Magnetization measurements were conducted on 
the La$_{1.64}$Eu$_{0.2}$Sr$_{0.16}$CuO$_{4}$ single crystals 
under {\it uniaxial} pressures up to 5 kbar applied along three 
independent directions, [001], [100] and [110]. 
A uniaxial pressure cell, which can be used in a commercial SQUID 
magnetometer (MPMS, Quantum Design), was developed. 
In order to reduce the background signal from the pressure cell, 
the main body is made of a high-strength plastic material. 
The background signal was found to be negligibly small, as compared 
with the Meissner signal of a few mg of the sample. 
The sample was pressurized uniaxially in the cell until it breaks up. 
On applying pressure, we sometimes observed a reduction of the magnitude 
of the superconducting signal. We believe that this is due to micro-cracks, 
which likely reduce the uniaxial nature and cause a broadening of 
the transition. In the following, therefore, we deal only with data, 
which did not show any appreciable reduction of the shielding signal. 

In this work on La$_{1.64}$Eu$_{0.2}$Sr$_{0.16}$CuO$_{4}$, we found 
for the first time anisotropy of the pressure effect on {\it T}$_{\rm c}$ 
within the CuO$_{2}$ plane. 
A gigantic uniaxial pressure effect was observed only for pressures 
applied along the tetragonal [110] direction, i.e. at 45 degrees 
to the stripe orientation. This demonstrates that stripes, 
the one-dimensional charge objects, indeed strongly couple to 
an anisotropic distortion of the lattice.

\begin{figure}[t]
\includegraphics[width=0.65\linewidth]{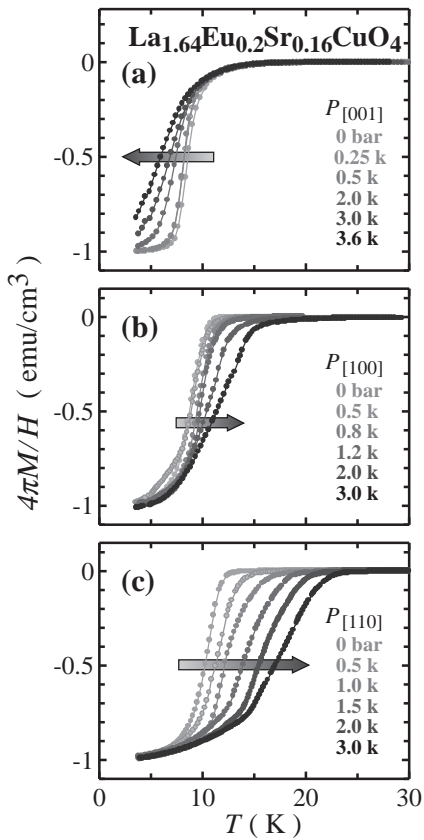}
\caption{Temperature dependence of magnetization for 
 La$_{1.64}$Eu$_{0.2}$Sr$_{0.16}$CuO$_{4}$ single crystals 
 under uniaxial pressures applied along (a) [001], (b) [100], and (c) [110]. 
 The measurements were made under a magnetic field of H= 10 Oe 
 by warming up the sample after zero field cooling.}
\label{fig2}
\end{figure}

The pressure dependence of the temperature dependent magnetization 
is summarized in Figure~\ref{fig2}. 
The data clearly demonstrate that the anisotropy of the uniaxial 
pressure effect exists not only between the in-plane and the out-of-plane 
directions but also within the tetragonal CuO$_{2}$ planes. 
As seen in Fig.~\ref{fig2}(a), the superconducting transition shifts 
to a lower temperature with increasing pressure applied along [001]. 
In remarkable contrast, {\it T}$_{\rm c}$ increases when the pressure 
is applied parallel to the CuO$_{2}$ planes (both the [100] and the [110] 
directions), as shown in Figs.~\ref{fig2}(b) and \ref{fig2}(c). 
This contrast between the in-plane pressure and the out-of-plane pressure 
has been known for many years. A substantial change of {\it T}$_{\rm c}$, 
associated with anisotropic pressure induced by epitaxial strain, 
was observed in underdoped La$_{2-x}$Sr(Ba)$_{x}$CuO$_{4}$ 
thin films~\cite{ref17,ref18}. 
In the case of films on LaSrAlO$_{4}$ substrate, a substantial 
increase of {\it T}$_{\rm c}$ up to 50 K was observed. 
The LaSrAlO$_{4}$ substrate has a slightly smaller in-plane lattice constant 
than those of bulk La$_{2-x}$Sr$_{x}$CuO$_{4}$, which results 
in the in-plane compression and the out-of-plane expansion of 
the La$_{2-x}$Sr$_{x}$CuO$_{4}$ film. 
The rapid increase of {\it T}$_{\rm c}$ has been ascribed to such 
an epitaxial strain effect. 
Consistent with this, a substantial decrease of {\it T}$_{\rm c}$ was 
observed for SrTiO$_{3}$ substrate with a large lattice constant 
which gives rise to the in-plane expansion and the out-of-plane compression. 
This strong dependence of {\it T}$_{\rm c}$ on anisotropic pressure was also 
directly demonstrated by a recent uniaxial pressure measurement~\cite{ref12}. 
Our observation of distinct differences between the in-plane and 
the out-of-plane pressure effects agrees well with these observations.

\begin{figure}[b]
\includegraphics[width=0.65\linewidth]{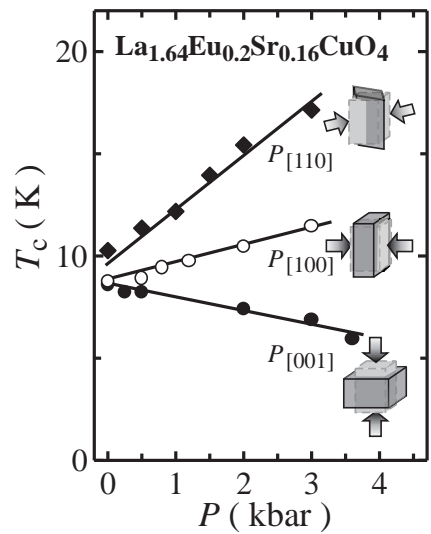}
\caption{Pressure dependence of {\it T}$_{\rm c}$ for 
 La$_{1.64}$Eu$_{0.2}$Sr$_{0.16}$CuO$_{4}$ with uniaxial pressures 
 applied along [100], [110] and [001]. {\it T}$_{\rm c}$ was defined 
 as a temperature where the diamagnetic signal reaches 50\% 
 of perfect diamagnetism in Fig. 2.}
\label{fig3}
\end{figure}

The most remarkable result here is the substantial difference between 
the [100] direction and the [110] direction, both within the CuO$_{2}$ plane. 
This is the first observation of an {\it in-plane} anisotropic pressure 
effect in high-{\it T}$_{\rm c}$ cuprates. 
Only when the pressure is applied along [110] do we see a strikingly 
large increase of {\it T}$_{\rm c}$ with increasing pressure. 
Indeed, the application of only several kbar causes {\it T}$_{\rm c}$ 
to almost double. 
In contrast, when the pressure is applied along [100], only a small 
change in {\it T}$_{\rm c}$ is observed. 
This implies that the in-plane pressure effect cannot be simply described 
by a superposition of pressure effects along two orthogonal directions 
but strongly depends on the symmetry of strains induced. 
This novel anisotropic pressure effect is visually summarized 
in Figure~\ref{fig3}. 
Note that the increase of {\it T}$_{\rm c}$ with {\it P}//[110] upon pressure, 
{\it dT$_{\rm c}$/dP}$_{[110]}$ $\sim$ 2.5 K/kbar is even larger 
than observed on La$_{2-x}$Ba$_{x}$CuO$_{4}$ around {\it x} $\sim$ 1/8 
with hydrostatic pressure~\cite{ref9,ref10}, previously 
the largest pressure effect ever observed in high-{\it T}$_{\rm c}$ cuprates.

We ascribe this striking in-plane anisotropy to the intimate coupling 
between the one-dimensional stripes and an anisotropic lattice distortion 
within the CuO$_{2}$ planes. 
The stripes are running along either [100] or [010], switching alternately 
from plane to plane along the c-axis~\cite{ref1}. 
Note that the [110] direction is rotated 45 degrees to the stripe directions. 
Since the symmetry of strain induced by the [110] pressure does not match 
with those of the one-dimensional stripes, it appears natural that 
the [110] pressure suppresses stripe formation and therefore enhances 
superconductivity. 
The [100] pressure, on the other hand, is parallel or perpendicular to 
the stripes, which gives rise to an orthorhombic strain matching 
with the local symmetry of stripes within each CuO$_{2}$ plane. 
The [100] strain therefore might not be so detrimental to stripe formation. 
In addition, the 90 degree rotation of the stripe direction from plane to 
plane may act to cancel out the [100] strain effects.

Such an intimate link between the symmetry of strain and the stripes 
may become even clearer by considering the presence of the LTT distortion, 
an alternate tilting of the CuO$_{6}$ octahedron along the stripe direction, 
that is known to stabilize stripe formation. 
In rare earth (RE = Nd, Eu, Sm) substituted La$_{2-x}$Sr$_{x}$CuO$_{4}$, 
the system switches from a stripe magnet to a superconductor upon hole doping 
at a composition away from the magic number 1/8. 
It has been proposed~\cite{ref19} that this critical phase boundary 
is determined by the strength of the LTT distortion, which can 
in turn be quantified by the tilting angle of the CuO$_{6}$ octahedra. 
The presence of a universal critical angle for switching was also 
demonstrated. The tilting angle decreases upon Sr doping and, eventually 
at the critical angle, drives the stripe magnet into SC. 
The [110] strain, at 45 degrees to the stripe direction [100] or [010], 
very likely reduces the LTT tilting angle because of the symmetry mismatch. 
This decrease of the tilting angle will shift the phase boundary 
between stripes formation and superconductivity to lower {\it x} 
and will substantially enhance {\it T}$_{\rm c}$ of the sample 
in the critical region. 
From these considerations, we believe that the extremely large uniaxial 
and anisotropic pressure effect within the CuO$_{2}$ planes 
in La$_{1.64}$Eu$_{0.2}$Sr$_{0.16}$CuO$_{4}$ is realized 
by an intimate interplay of the strain-stripe coupling mediated 
by the lattice distortion and the stripe criticality. 

The remarkable uniaxial pressure effect observed in this study should 
help to clarify the longstanding issue of the interplay 
between stripes and high temperature superconductivity and, as a result, 
might shed new light on the high-{\it T}$_{\rm c}$ problem. 
First, by checking for this anisotropy, we may identify the contribution 
from the stripes and particularly dynamical stripes to superconductivity 
itself. Indeed, uniaxial pressure measurements are now under progress 
on samples with Sr content {\it x} away from the critical composition 
{\it x} = 0.16, in order to explore the influence of dynamical stripes. 
Second, using [110] pressure, we should be able to switch the stripe magnet 
to a superconductor using relatively low pressures of a few kbars, 
if the sample is located at the stripe side of the critical region. 
We indeed observed such a stripe magnet to superconductor transition at 1 kbar 
in a {\it nonsuperconducting} La$_{1.65}$Eu$_{0.2}$Sr$_{0.15}$CuO$_{4}$, 
which has a marginally smaller Sr content than the samples shown 
in Figures~\ref{fig1}-\ref{fig3}. 
This magnitude of pressure can be generated quite easily without 
any special facility. 
Besides, in contrast with the hydrostatic pressure, the sample 
in the uniaxial experiments is covered by the piston cylinder only 
from the top and bottom and beam probes can reach the sample quite easily. 
It could be possible therefore to carry out a variety of physical 
measurements, including angle resolved photoemission spectroscopy (ARPES), 
neutron diffraction etc., to trace the evolution from stripe magnet to 
superconductor continuously using the same piece of crystal. 

In summary, we have discovered a substantially large {\it in-plane} 
anisotropic uniaxial pressure effect on {\it T}$_{\rm c}$ 
in La$_{1.64}$Eu$_{0.2}$Sr$_{0.16}$CuO$_{4}$, located 
at the critical vicinity to the (static) stripe-superconductivity boundary. 
This result implies very strong coupling between the stripe formation 
and the anisotropic lattice distortion. 
The charge ordered state, a self organization of electrons, has 
a lower symmetry than that of the original lattice and is stabilized 
by coupling to the lattice. In this sense, what we have observed 
in La$_{1.64}$Eu$_{0.2}$Sr$_{0.16}$CuO$_{4}$ should be universal 
to charge ordered systems in general. 
Uniaxial pressure has proved itself as an extremely useful probe to 
explore novel strongly correlated electron systems.\par

The author would like to thank K. Kitazawa for discussion and his continuous encouragement and N. E. Hussey for discussion and critical reading of manuscript. This work is partly supported by a Grant-in-Aid for Scientific Research from the Ministry of Education, Culture, Sports, Science and Technology, Japan and New Energy and Industrial Technology Development Organization (NEDO), Japan.\par



\begin{thebibliography}{00}
\bibitem{ref1} 
 J. M. Tranquada, B. J. Sternlieb, J. D. Axe, Y. Nakamura, and S. Uchida, 
 Nature {\bf 375}, 561 (1995). 
\bibitem{ref2} 
 A. R. Moodenbaugh, Youwen Xu, M. Suenaga, T. J. Folkerts, and R. N. Shelton, 
 Phys. Rev. B {\bf 38}, 4596 (1988). 
\bibitem{ref3} 
 J. D. Axe, A. H. Moudden, D. Hohlwein, D. E. Cox, K. M. Mohanty, 
 A. R. Moodenbaugh, and Youwen Xu, 
 Phys. Rev. Lett. {\bf 62}, 2751 (1989).
\bibitem{ref4} 
 H. A. Mook, Pengcheng Dai, F. Do\u{g}an, and R. D. Hunt, 
 Nature {\bf 404}, 729 (2000).
\bibitem{ref5} 
 P. Bourges, Y. Sidis, H. F. Fong, L. P. Regnault, J. Bossy, 
 A. Ivanov, and B. Keimer, 
 Science {\bf 288}, 1234 (2000).
\bibitem{ref6} 
 J. Zaanen, Science {\bf 286}, 251 (1999).
\bibitem{ref7} 
 S. A. Kivelson, E. Fradkin, and V. J. Emery, 
 Nature {\bf 393}, 550 (1998).
\bibitem{ref8} 
 C. Di Castro, L. Benfatto, S. Caprara, C. Castellani, and M. Grilli, 
 Physica (Amsterdam) {\bf 341-348C}, 1715 (2000). 
\bibitem{ref9} 
 S. Katano, S. Funahashi, N. M\^{o}ri, Y. Ueda, and J. A. Fernandez-Baca, 
 Phys. Rev. B {\bf 48}, 6569 (1993). 
\bibitem{ref10} 
 M. Ido, N. Yamada, M. Oda, Y. Segawa, N. Momono, A. Onodera, 
 Y. Okajima, and K. Yamaya, 
 Physica (Amsterdam) {\bf 185-189C}, 911 (1991). 
\bibitem{ref11} 
 B. Lake, H. M. Ronnow, N. B. Christensen, G. Aeppli, K. Lefmann, 
 D. F. McMorrow, P. Vorderwisch, P. Smeibidl, N. Mangkorntong, 
 T. Sasagawa, M. Nohara, H. Takagi, and T. E. Mason, 
 Nature {\bf 415}, 299 (2002). 
\bibitem{ref12} 
 S. Arumugam, N. M\^{o}ri, N. Takeshita, H. Takashima, T. Noda, 
 H. Eisaki, and S. Uchida, 
 Phys. Rev. Lett. {\bf 88}, 247001 (2002). 
\bibitem{ref13} 
 M. K. Crawford, R. L. Harlow, E. M. McCarron, W. E. Farneth, 
 J. D. Axe, H. Chou, and Q. Huang, 
 Phys. Rev. B {\bf 44}, 7749 (1991). 
\bibitem{ref14} 
 H. -H. Klauss, W. Wagener, M. Hillberg, W. Kopmann, H. Walf, 
 F. J. Litterst, M. H\"{u}cker, and B. B\"{u}chner, 
 Phys. Rev, Lett. {\bf 85}, 4590(2000). 
\bibitem{ref15a} 
 I. Tanaka and H. Kojima, 
 Nature {\bf 337}, 21 (1989). 
\bibitem{ref15b} 
 T. Sasagawa, K. Kishio, Y. Togawa, J. Shimoyama, and K. Kitazawa, 
 Phys. Rev. Lett. {\bf 80}, 4297 (1998). 
\bibitem{ref16} 
 T. Noda, H. Eisaki, and S. Uchida, 
 Science {\bf 286}, 265 (1999).
\bibitem{ref17} 
 H. Sato and M. Naito, 
 Physica (Amsterdam) {\bf 274C}, 221 (1997). 
\bibitem{ref18} 
 J. -P. Locquet, J. Perret, J. Fompeyrine, E. Machler, J. W. Seo, 
 and G. Van Tendeloo, Nature {\bf 394}, 453 (1998). 
\bibitem{ref19} 
 B. B\"{u}chner, M. Breuer, A. Freimuth, and A. P. Kampf, 
 Phys. Rev. Lett. {\bf 73}, 1841 (1994). 
\end{thebibliography}
\end{document}